\documentclass[letter]{jpsj3}

\title{Brownian dynamics around the core of self-gravitating systems}

\author{Tohru Tashiro\thanks{E-mail address: tashiro@cosmos.phys.ocha.ac.jp}
and Takayuki Tatekawa$^{1}$
}
\inst{Department of Physics, Ochanomizu University, 2-1-1 Ohtuka, Bunkyo, Tokyo 112-8610, Japan \\
$^{1}$Center for Computational Science and e-Systems, Japan Atomic Energy Agency, 6-9-3 Higashi-Ueno, Taito, Tokyo, 110-0015, Japan
}

\abst{We derive the non-Maxwellian distribution of self-gravitating $N$-body systems around the core by a model based on the random process with the additive and the multiplicative noise. The number density can be obtained through the steady state solution of the Fokker-Planck equation corresponding to the random process. We exhibit that the number density becomes equal to that of the King model around the core by adjusting the friction coefficient and the intensity of the multiplicative noise. We also show that our model can be applied in the system which has a heavier particle. Moreover, we confirm the validity of our model by comparing with our numerical simulation.}

\kword{Brownian motion, Langevin equation, multiplicative noise, Fokker-Planck equation, self-gravitating system, globular clusters}

\begin{document}
\maketitle

How relevant is the equilibrium statistical mechanics when we describe the steady state of a self-gravitating system (SGS) where many particles interact via the gravitational force?
Let's assume that the state of the SGS with equal mass $m$ becomes isothermal with temperature $T$ and the particles of the system will be distributed spherically symmetrically. Then, the structure in the phase space can be determined by the Maxwell-Boltzmann distribution. For example, the number density at a radial distance $r$ in the real space is
\begin{equation}
n_{\rm MB}(r) \propto e^{-\frac{m}{k_{\rm B}T}\Phi(r)} \ ,
\label{nMB}
\end{equation}
where $\Phi(r)$ is the mean gravitational potential per mass generated by this whole system at $r$ and $k_{\rm B}$ is the Boltzmann constant.
This potential should satisfy a relation with the number density by the Poisson equation $\bigtriangleup\Phi(r)=4\pi Gmn_{\rm MB}(r)$
where $G$ is the gravitational constant.
A special solution of eq.(\ref{nMB}) and this Poisson equation is $n_{\rm MB}(r)=k_{\rm B}T/2\pi Gm^2r^2$ known as the singular isothermal sphere \cite{Binney87}. This solution has two problems: infinite density at $r=0$ and infinite total mass. Even though we solve the equations with finite density at $r=0$, the solutions behave $\propto r^{-2}$ at large $r$, and so we cannot get around the infinite total mass problem. In either case, the solutions are unrealistic.
As expected, it is unreasonable to apply the {\em standard} statistical mechanics to SGS where the interaction is a long-range force.

However, the real systems in the universe, e.g. globular clusters, galaxies etc., have various structures. As for the most of globular clusters, it is known that the number densities of them in the real space have a flat core and behave as a power law outside the core. 
King interpreted these profiles by introducing the new distribution function (DF) in the phase space, known as the {\em lowered Maxwellian}, which becomes zero when the total energy is greater than a certain value by subtracting a constant from the original Maxwell-Boltzmann distribution. This is called as the King model \cite{King1966}.
The approximation of the number density of the King model, $n_{\rm KM}({r})$, around the core is
\begin{equation}
n_{\rm KM}({r}) \propto \frac{1}{(1+r^2/a^2)^{3/2}} \ ,
\label{KM}
\end{equation}
where $a$ is the core radius.
Since he put forward this model,
the number density has been applied to
fitting for the surface brightness of many globular
clusters, for example as in ref.\cite{Kingexp}.

In this Letter, we try to derive this non-Maxwellian distribution in the real space around the core from a new point of view. Our simple model uses a Brownian dynamics described by random process with the additive and the multiplicative noise, which is quite different from the King model because his procedure was done to the DF in the steady state. 
Physically, this model represents that the gravitational force induced by $\Phi(r)$, the {mean force}, fluctuates in time.
We get the Fokker-Planck equation from the Langevin equation and show that the same result as the King model can be obtained from the steady state solution.
In addition, we show that our model can deal with the case that the system includes {\em another particle} whose mass is $M (> m)$, corresponding to the black hole in a globular cluster.

%\section{Steady number density in $N$-body simulation}
%\label{simu}

Now, we investigate the steady number density (SND) of the SGS with mass $m$ including a particle with mass $M$ by using $N$-body simulation. Especially, we show the results that the density profiles in the steady state have a core and behave as a power law.
The system is composed of $N=10000$ particles. At $t=0$, all velocities of the particles are zero and they are distributed by $n_0(r) \propto {(1+r^2/{a_p}^2)^{-5/2}} ~(0\le r \le 4{a_p})$ which is the density in the real space of Plummer's solution \cite{Binney87}. In this SGS we put {\em another particle} with mass $M$ in the origin at $t=0$. We shall change the mass as $M/m=1$, $5$, and $10$. Throughout this Letter, we adopt a unit system where the core radius of the Plummer's solution $a_p$, initial free fall time $t_{ff}$, and the total mass $N\cdot m$ are unity.

We started the $N$-body simulation under these conditions.
For dynamical evolution, we use GRAPE-7, special purpose computer
for gravitational force \cite{Kawai2006}.
For computation of the gravitational force, we apply
Plummer's softening: the potential energy between the $i$th and the $j$th particle separated by a distance $r_{ij}$ is 
$-{G m^2}/{\sqrt{{r_{ij}}^2+{\varepsilon_{s}}^2}}$
where $\varepsilon_{s}$ is the softening parameter. We set
$\varepsilon_{s} = 10^{-3}$.
For evolution, we use sixth order symplectic integrator \cite{Yoshida1990}.
The time step for the simulations is defined as $\Delta t = 10^{-5}$.
We carried out the simulations until $t=100~t_{ff}$. During simulations,
the error of the total energy is less than $0.1 \%$.

\begin{figure}[h]
\begin{center}
 \includegraphics[scale=.5]{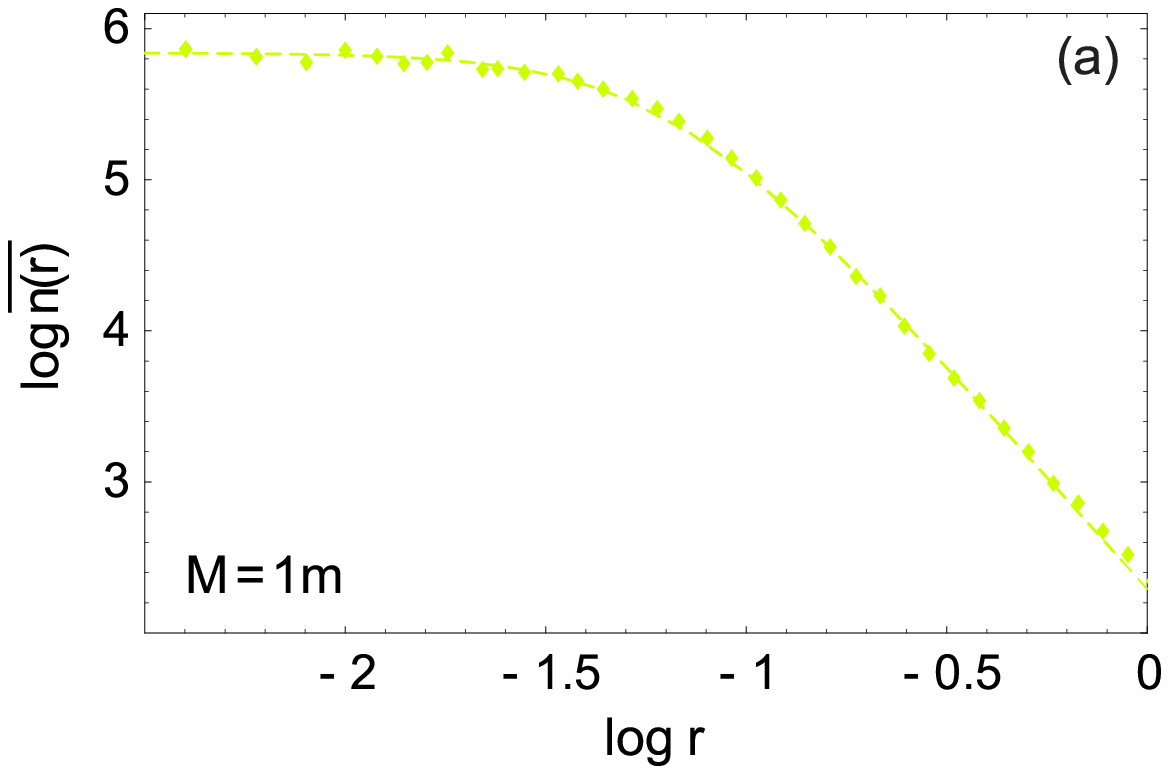}

 \includegraphics[scale=.5]{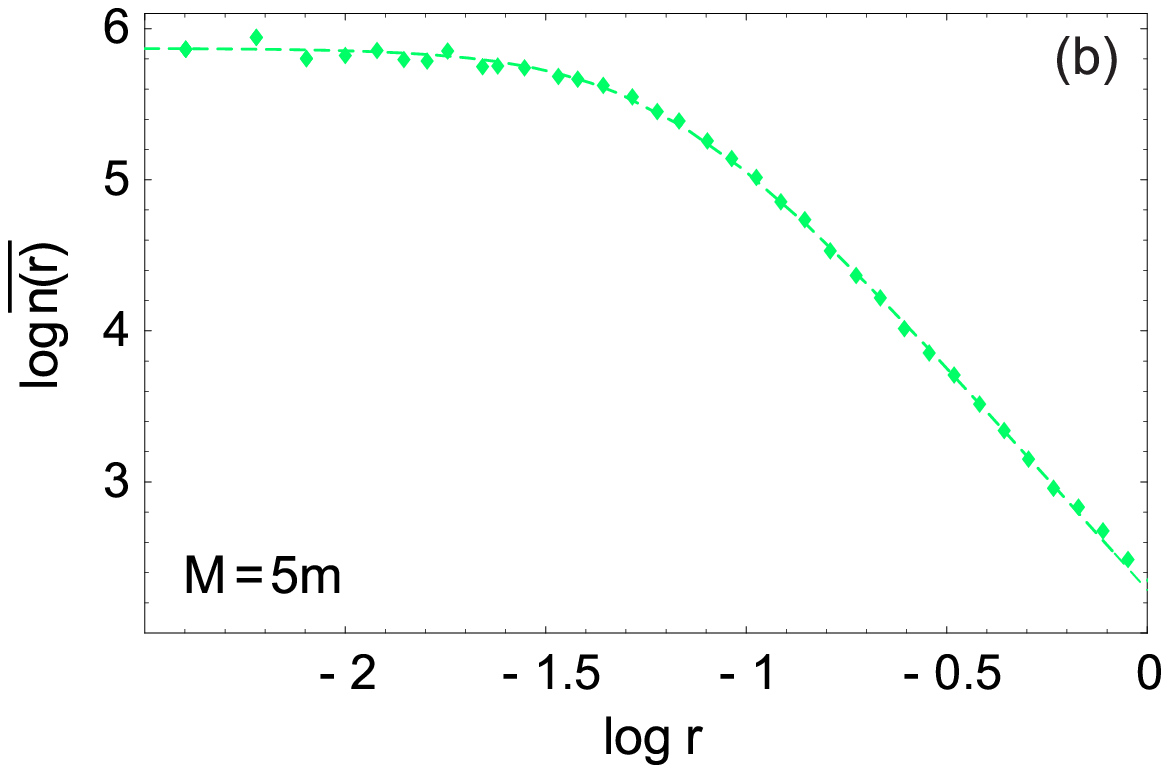}

 \includegraphics[scale=.5]{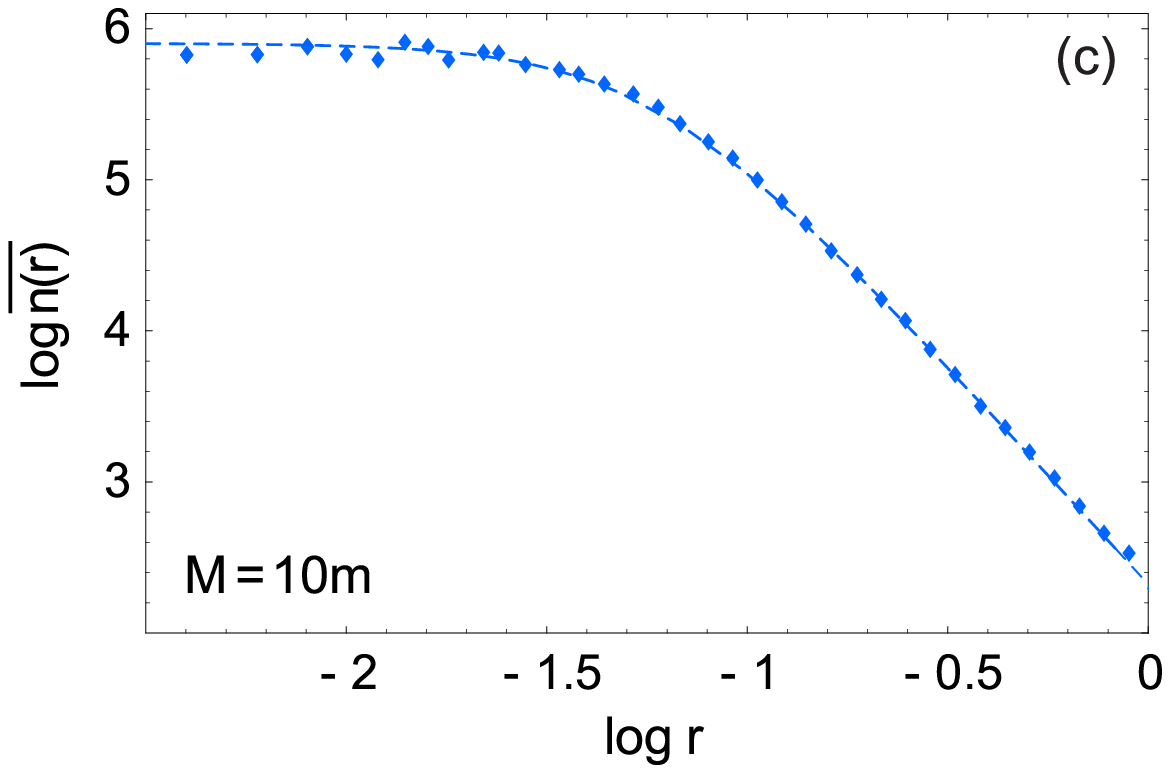}
 \caption{(Color online) Logarithm of the steady number density $\overline{n(r)}$ as a function of $\log r$ for (a) $M/m=1$, (b) $M/m=5$, and (c) $M/m=10$. In each figure, the dashed curve and symbols denote a fitting curve $\overline{n_{\rm fit}(r)}=C/(1+r^2/a^2)^\beta$ and the result of our numerical simulation, respectively.}
 \label{fig1}
\end{center}
\end{figure}

At first, most of the particles collapse into the origin within several $t_{ff}$. Approximately after $20~t_{ff}$, the distribution becomes stable and the system goes to the steady state.
In Fig.\ref{fig1}, we show the logarithm of SND as a function of $\log r$ for $M/m=1$, $5$, and $10$. For each $M$, the SND has a core and behaves as a power low at larger $r$ than the core radius.

We now fit SNDs around the core by
$\overline{n_{\rm fit}(r)}=C/(1+r^2/a^2)^\beta$.
The results are summarized in Table \ref{tabpara}.
For $M/m=1$ and 5, $\beta\simeq3/2$ which is similar to the exponent of the King model.
The density at the origin $C$ increases as $M$ is increased, which is simply
understood as a result that many particles are attracted by the heavier particle.

\begin{table}[h]
  \renewcommand{\arraystretch}{1.52}
  \caption{The best fitting parameters of a function $C/(1+r^2/a^2)^\beta$ for steady number densities shown in Fig.\ref{fig1}}
  \label{tabpara}
  \begin{center}
    \begin{tabular}{cp{0.3cm}cp{0.3cm}cp{0.3cm}c}
      %\hline
      \hline
      %      & & &  \\
      $M/m$ & & $a$ & & $\beta$ & & $C$ \\
      %      & & &  \\
      \hline
      %    & & &  \\
      1 &  & $6.44\times10^{-2}$  &   & $1.49$ &  & $6.91\times10^5$        \\
      %    & & &  \\
      5 &  & $6.28\times10^{-2}$  &   & $1.49$ &  & $7.41\times10^5$        \\
      %    & & &  \\
      10&  & $5.84\times10^{-2}$  &   & $1.45$ &  & $7.99\times10^5$        \\
      %    & & &  \\
      %\hline
      \hline
    \end{tabular}
  \end{center}
\end{table}

%\section{Simple model}
%\label{tm}

In order to explain these results and derive the non-Maxwellian distribution around the core, we demonstrate the simple model based on Brownian motion which is quite different with the King model. The reason why Brownian motion appears in the SGS is as follows. After the collapse, the density around the core becomes high. Thus the particles around there disturb the orbits of others repeatedly, so that the movements become random \cite{com3}. As the time to occur this disturbance, we introduce the local two-body relaxation time $t_{\rm rel}$ \cite{Spitzer1987}:
\begin{equation}
t_{\rm rel}(r) = \frac{0.065\sigma(r)^{3}}{G^2\overline{n(r)}m^2\ln(1/\varepsilon_{s})} \ ,
\end{equation}
where $\sigma(r)$ is the standard deviation of the velocity at $r$ and we adopted $\ln(1/\varepsilon_{s})$ as the Coulomb logarithm.

\begin{figure}[h]
\begin{center}
 \includegraphics[scale=.55]{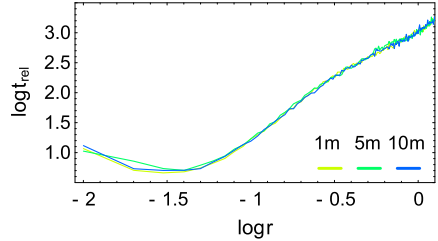}
 \caption{(Color online) Logarithm of the local two-body relaxation time $t_{\rm rel}$ as a function of $\log r$ for $M/m=1$, 5, and 10.}
 \label{fig2}
\end{center}
\end{figure}

Figure~\ref{fig2} shows the logarithm of $t_{\rm rel}$, which is calculated by using $\sigma(r)$ and $\overline{n(r)}$ obtained from our $N$-body simulation, as a function of $\log r$. As expected, $t_{\rm rel}$ around the core is short. Our simulation continues after the collapse during about $80~t_{ff}$ which is sufficiently longer than $t_{\rm rel}$ around the core. As the radius increases, however, $t_{\rm rel}$ becomes longer than the rest of our simulation time, which means that the Brownian motion does not occur at large $r$.
Therefore, note that our model is valid only in the neighborhood of the core.

When constructing our model, the following points are premised: the model describes the dynamics near the steady state and the mean distribution is spherically symmetric. 
As well known, a gravitational force at $r$ arising from such a spherically symmetric system depends only on the particles which exist inside the sphere with radius $r$ and this attractive force acts along the radial direction. In other words, this mean force $-f(r)$ is the gradient of the mean potential: $-f(r)=-m{\partial\Phi(r)}/{\partial r}$.
Of course, $\lim_{r\rightarrow0}f(r)=0$. Hence, we assume
\begin{equation}
f(r)=\alpha m^2r+O(r^2) \ ,
\label{f(r)}
\end{equation}
where $\alpha\equiv 4\pi GC/3$.
This condition is necessary in order that the SND has a core as will become clear later. Conversely, if the SND has a core, $f(r)$ is represented as eq.(\ref{f(r)}), which is easily understood as following way. Now, the SND is almost constant in the neighborhood of the core, and so the sum of the mass of particles which exist inside small $r$ is $M(r)\sim 4\pi Cmr^3/3$. Therefore, $f(r)=GmM(r)/r^2\sim  4\pi GCm^2r/3$. After all, the SND is linked to the mean force $f(r)$ self-consistently around the core.

For the case $M/m=1$, we can identify {\em another particle} with others. Contrary to this, we must consider the effect of the particle for the case $M/m\neq1$. Now, we suppose that the heavier particle exists at the origin. Then, the attractive force by this particle at $r$ is $-F(r) = -GmM/r^2$. We can estimate $f(r)$ around there as $f(r)\sim4\pi Gm^2Cr/3$. Thus, ${F(r)}/{f(r)}\sim{3Mr^{-3}}/{4\pi Cm}$. This ratio has a meaning when $r\sim10^{-2}$. Therefore, if $r$ is smaller than the radius, particles are influenced by not only $f(r)$ but also $F(r)$, so that the core should disappear. In fact, we have done a numerical simulation with the heavier particle fixed at the origin, where this result is confirmed. On the other hand, Miocchi improved the King model in order to describe steady state of a globular cluster including an intermediate-mass black hole and reported that the density becomes cuspy as the mass of the black hole increases \cite{Miocchi2007}.
Because the heavier particle of our numerical simulation is not heavy so much, the particle is not trapped at the origin. Therefore, we do not consider the effect of the heavier particle explicitly and we suppose that the particle influence SGS through the density at the origin $C$: as $M$ becomes larger, it attracts more particles and $C$ increases as shown in Table \ref{tabpara}. Thus, $\alpha=4\pi GC/3$ is increasing function of $M$.

It is natural to consider that the distribution fluctuates around the mean value because of the many disturbances. The fluctuating part of distribution should not be spherically symmetric, so that this produces the forces along not only the radial direction, but also the other directions. We assume that they are random forces and set the intensity of them at $r$ $2E(r)^2$. In addition to such random forces resulting from the fluctuating distribution, a particle at $r$ is expected to be influenced by random forces generated from neighbor particles. We set the intensity $2D$ which is independent with position.

Brownian motion with the above assumptions is described by the following Langevin equations in spherical coordinates: the radial direction
\begin{equation}
m\gamma\dot{r} = -f(r) + \sqrt{2}E(r)\eta_r(t) + \sqrt{2D}\xi_r(t) \ , \label{rad}
\end{equation}
the elevation direction
\begin{equation}
m\gamma r\dot{\theta} = \sqrt{2}E(r)\eta_\theta(t) + \sqrt{2D}\xi_\theta(t) \ , \label{ele}
\end{equation}
and the azimuth direction
\begin{equation}
m\gamma r\sin\theta\dot{\phi} = \sqrt{2}E(r)\eta_\phi(t) + \sqrt{2D}\xi_\phi(t) \label{azi} \ ,
\end{equation}
where $\gamma$ is the coefficient of dynamical friction in the low velocity limit and independent on the velocity \cite{Binney87}. In the Chandrasekhar dynamical friction formula, the coefficient is more complicated \cite{Binney87,Chandra1943}. But we use the coefficient in such a limit, because the density around the core is so high that particles around there move slowly.

Since we consider the Brownian dynamics near the steady state, the inertial terms are neglected \cite{foot}. 
The noises in each Langevin equation, $\xi$ and $\eta$, are zero-mean white Gaussian and are correlated only to themselves.
Of course, the correlation function is the Dirac delta function  \cite{footL}.

Here, we make the following final assumption: the intensity $E(r)$ behaves similarly to $f(r)$ around the core, that is $E(r)=\sqrt{\epsilon}f(r)=\sqrt{\epsilon}\alpha m^2r+O(r^2)$ where $\epsilon$ is a positive constant.
Then, the first and the second term on the right-hand side of eq.(\ref{rad}) become
\begin{equation}
-f(r)\{1-\sqrt{2\epsilon}\eta_r(t)\} \ .
\label{fluMF}
\end{equation}
This represents that the mean force fluctuates. As we wrote before, the distribution in the near steady state is expected to fluctuate around the mean value which yields the mean force $-f(r)$, so that the total gravitational force from the whole system along the radial direction is described as in eq.(\ref{fluMF}).

We have the Fokker-Planck equation governing the spherically symmetric probability distribution function (PDF) $P(r,t)$
\begin{eqnarray}
\frac{\partial}{\partial t}P(r,t) &=& \frac{D}{(m\gamma)^2}\left\{\frac{\partial^2}{\partial r^2}+\frac{2}{r}\frac{\partial}{\partial r}\right\}P(r,t) + \frac{1}{m\gamma}\frac{1}{r^2}\frac{\partial}{\partial r}r^2f(r)P(r,t) \nonumber \\
 && \hspace{-1.6cm}\mbox{}+\frac{\epsilon}{(m\gamma)^2}\left\{\frac{\partial^2}{\partial r^2}f(r)^2+\frac{2}{r}\frac{\partial}{\partial r}f(r)^2-\frac{1}{2r^2}\frac{\partial}{\partial r}r^2(f(r)^2)'\right\}P(r,t) \ , \nonumber \\
\label{PDEQ}
\end{eqnarray}
where the prime indicates derivative with respect to $r$.
Then, the PDF with the Jacobian $\rho(r,t)\equiv4\pi r^2P(r,t)$ satisfies the following Fokker-Planck equation.
\begin{eqnarray}
\frac{\partial}{\partial t}\rho(r,t) &=& \frac{D}{(m\gamma)^2}
\left\{\frac{\partial^2}{\partial r^2}-\frac{\partial}{\partial r}\frac{2}{r}\right\}\rho(r,t) + \frac{1}{m\gamma}\frac{\partial}{\partial r}f(r)\rho(r,t) \nonumber \\
 && \hspace{-1.1cm} \mbox{}+\frac{\epsilon}{(m\gamma)^2}\left\{\frac{\partial}{\partial r}f(r)\frac{\partial}{\partial r}f(r)-\frac{\partial}{\partial r}\frac{2}{r}f(r)^2\right\}\rho(r,t)
\label{rhoDEQ}
\end{eqnarray}
This equation is useful when integrating with respect to $r$.

The steady state solution $\rho_{\rm st}(r)$ satisfies the equation~(\ref{rhoDEQ}) with the left-hand side zero. By integrating the equation with respect to $r$, we have
\begin{eqnarray}
&&\left\{\frac{D}{(m\gamma)^2}+\frac{\epsilon}{(m\gamma)^2}f(r)^2\right\}\rho_{\rm st}'(r) \nonumber \\
&&\hspace{-.4cm}-\left[\frac{D}{(m\gamma)^2}\frac{2}{r}
 - \frac{\epsilon}{(m\gamma)^2}\left\{f(r)f'(r)-\frac{2}{r}{f(r)^2}\right\}  - \frac{f(r)}{m\gamma}\right]\rho_{\rm st}(r)  \nonumber \\
&=& \mbox{const.}  \ .
\label{rhoDEQ2}
\end{eqnarray}

Now, we impose the binary condition that $P_{\rm st}(r)\equiv\rho_{\rm st}(r)/(4\pi r^2)$ and the derivative do not diverge at the origin. Then, when $r\rightarrow0$, $\rho_{\rm st}(r)=O(r^2)$ and $\lim_{r\rightarrow0}\rho_{\rm st}'(r)=\lim_{r\rightarrow0}4\pi(2rP_{\rm st}(r)+r^2P'_{\rm st}(r))=0$ by which the constant on the right-hand side of eq.(\ref{rhoDEQ2}) is decided and we obtain
\begin{equation}
\rho_{\rm st}'(r)=
-\frac
{r{f(r)}\left\{{\epsilon}f'(r)+m\gamma\right\}-{2}\left\{{D} + {\epsilon}f(r)^2\right\}}
{r\left\{{D}+{\epsilon}f(r)^2\right\}}\rho_{\rm st}(r) \ .
\end{equation}

Let's represent $f(r)$ by the first order of $r$. Then,
\begin{equation}
\rho_{\rm st}'(r)=
-\frac
{-{2}\frac{D}{\epsilon\alpha^2m^4}+\left\{\frac{\gamma}{\epsilon m\alpha}-1\right\}r^2}
{r\left\{\frac{D}{\epsilon\alpha^2m^4}+r^2\right\}}\rho_{\rm st}(r) \ .
\label{lastDEQ}
\end{equation}
Here, if we set
\begin{equation}
a^2\equiv\frac{D}{\epsilon\alpha^2m^4} \ \  \mbox{and} \ \ \beta\equiv\frac{1}{2}\left(\frac{\gamma}{\epsilon m\alpha}+1\right) \ ,
\label{para}
\end{equation}
the equation (\ref{lastDEQ}) can be solved like
\begin{equation}
\rho_{\rm st}(r) \propto \frac{r^2}{(1+r^2/a^2)^\beta} \ ,
\end{equation}
which yields
\begin{equation}
P_{\rm st}(r) \propto \frac{1}{(1+r^2/a^2)^\beta} \ .
\label{main}
\end{equation}
Since the stochastic behavior of a particle of the SGS is governed by this PDF, the SND of this system can be obtained by multiplying this equation by a constant.

%\section{Discussion}
%\label{dis}

%In this section, we investigate the results derived in the preceding section and understand the roles of two noises and the heavier particle in eq.(\ref{main}). Additionally, we discuss the difference between the King model and our model.

As in eq.(\ref{para}), the exponent $\beta$ must be larger than $1/2$, which does not contradict our numerical simulation shown in Table \ref{tabpara}.
In order that the equation (\ref{main}) corresponds  completely to the King model, $\beta=3/2$ or $\gamma = 2\epsilon m\alpha$ must hold.
We can regard this relation between the friction coefficient $\gamma$ and the intensity of the multiplicative noise $\epsilon$ as a kind of {\em fluctuation-dissipation relation} \cite{Kubo}, which usually plays an important role when a random process with an additive noise goes to the equilibrium state described by the Maxwell-Boltzmann distribution.

The core radius $a$ is proportional to a square root of the intensity of the additive noise $D$ owing to eq.(\ref{para}). Then, the intensity spreads the region where the density is almost constant. This is recognized as the effect of this noise which has the property that it makes a system homogeneous.

Now, let's examine the role of the mass of the heavier particle $M$ in this system by the naive discussion. As written previously, $\alpha$ is increasing function of $M$. $\alpha$ exists in the denominator of $a$ and $\beta$. Then, both values should be reduced when $M$ is increased if other parameters are independent on $M$. These theoretical expectations are consistent with our numerical results shown in Table \ref{tabpara}.

How the result eq.(\ref{main}) changes if the mean force does not fluctuate? The steady state solution of eq.(\ref{PDEQ}) with $\epsilon=0$ is
\begin{equation}
P_{\rm st}(r) \propto e^{-\frac{m^2\gamma}{D}\Phi(r)} \ .
\label{MB2}
\end{equation}
Therefore, our result goes to the singular isothermal sphere as discussed at the beginning of this Letter when the mean force does not fluctuate.

Here, we examine the relation between the King model and our model. King transformed the DF in the phase space in order to avoid the singular isothermal sphere. In our model, we introduce the multiplicative noise into the system influenced by the mean force and the additive noise whose PDF becomes Maxwellian in the steady state as shown in eq.(\ref{MB2}), so that the non-Maxwellian DF eq.(\ref{main}) is derived. In short, although these procedures seem to be different, they may have the same meaning at least around the core.

%\section{Concluding remarks}
%\label{cr}

In conclusion, we have derived the non-Maxwellian DF eq.(\ref{main}) by the Brownian dynamics with the fluctuating mean force and the additive white noise. The number density derived from the DF can be the same as that of the King model around the core by controlling the friction coefficient and the intensity of the multiplicative noise. Furthermore, our model can be valid in the SGS with a heavier particle.
Of course, these results are consistent with our numerical simulation.
We can say that such a stochastic dynamics occurs behind the background of the King model. Finally, note that our result is available only in the neighborhood of the core. So, we must derive the density globally by further extended model and investigate the difference between the model and the King model, in the near future.

\begin{acknowledgements}

We would like to thank Prof. Masahiro Morikawa, Dr. Osamu Iguchi, and the members of Morikawa lab. for the
extensive discussions. 
All numerical computations were carried out on GRAPE system at Center for Computational Astrophysics, CfCA, of National Astronomical Observatory of Japan.

\end{acknowledgements}

%\appendix


\begin{thebibliography}{l}
\bibitem{footL}
It may not be natural that correlations of the random forces generated from the gravity are described by the Dirac delta function. But in this letter, for simplicity, the correlation times are assumed to be negligible. In other words, the time resolution of our simple model in the over-damped limit is assumed to be much longer than the correlation times.
\end{thebibliography}
\end{document}